\documentclass{ws-rv975x65}
\usepackage{subfigure}     % required only when side-by-side / subfigures are used
\usepackage{ws-rv-thm}     % comment this line when `amsthm / theorem / ntheorem` package is used
\usepackage{ws-rv-van}     % numbered citation/references (default)
\usepackage{ws-index}     % to produce multiple indexes
\makeindex
\newindex{aindx}{adx}{and}{Author Index}       % author index
\renewindex{default}{idx}{ind}{Subject Index}  % subject index

\usepackage{units}
\usepackage{xspace}
\usepackage{lineno}
\begin{document}

\chapter[Observations of a diffuse flux of cosmic neutrinos]{Observations of diffuse fluxes of cosmic neutrinos}\label{diffuse_cw}

\author[C. Wiebusch]{Christopher H. Wiebusch}
\index[aindx]{Wiebusch, C.} % or \aindx{Author, F.}

\address{RWTH Aachen University, III.Physikalisches Institut,\\
Otto Blumenthal Strasse, 52074 Aachen, Germany, \\
wiebusch@physik.rwth-aachen.de%\footnote{Affiliation footnote.}
}

\begin{abstract}

In this contribution the current observational results for
the diffuse flux of high-energy astrophysical neutrino flux are reviewed. 
In order to understand the science implications, the measurements in different   detection channels are discussed and 
results are compared. The discussion focusses is the energy spectrum, the 
flavor ratio and large scale anisotropy.

%% \emph{We would like to invite you to write an article for a review volume, Neutrino Astronomy--Current status, future prospects.  The book, to be published by World Scientific, will be divided into four chapters, each including several articles as indicated in the attached outline.  We hope you will contribute an article of about 12 pages on observations of diffuse neutrinos with IceCube.
%% The contract with World Scientific calls for submission of the files by August 1, 2015.  We would therefore like to receive your contribution by mid-June.   The attached outline includes the url for the style files.}

\end{abstract}
%\markright{Customized Running Head for Odd Page} % default is Chapter Title.
\body

\section{Introduction}

For a long time, the detection of high-energy cosmic neutrinos as 
cosmic messengers  has been an outstanding goal of astroparticle physics.
Their observation has been proposed %shortly after their discovery 
by Markov\cite{Markov:1960vja} already in the 60th of last century.
The proposed method was the detection of up-going muons as signature of 
a charged-current (CC) muon neutrino interaction below the detector. 
Based on this signature atmospheric neutrinos were discovered
 in deep underground detectors\cite{Achar:1965ova}\cite{Reines:1965qk}.
Soon it was realized that the expected astrophysical fluxes are be
small and  cubic-kilometer sized detectors would be
needed to accomplish the goal\cite{Gaisser:1994yf}. A key concept 
became the instrumentation of optically transparent natural
media with photo-sensors to construct large Cherenkov detectors.
A major step was achieved by the BAIKAL 
collaboration\cite{Belolaptikov:1997ry}, which first succeeded 
to install and operate a large volume Cherenkov detector using the deep water 
of lake Baikal. Thiss effort was rewarded by the first observation of 
atmospheric 
neutrino events in an open natural environment\cite{Balkanov:1997gw},
see also contribution by V. Aynutdinov.
Shortly after, the AMANDA neutrino telescope
successfully demonstrated the feasibility of the construction and operation
of a large Cherenkov detector in glacier ice\cite{Andres:1999hm}. It was the first neutrino telescope to observe high-energy atmospheric neutrinos in larger quantity\cite{Andres:2001ty} and to exclude 
optimistic astrophysical models\cite{Achterberg:2007qp}. In parallel to these 
efforts  neutrino telescopes in deep oceans have also been 
brought into operation.
The ANTARES neutrino telescope\cite{antares:2011nsa}, see also contribution by P. Coyle, increased the effective area with respect to AMANDA and has been operational since 2006. In all these experiments no
indications of cosmic neutrinos have been found\cite{Aguilar:2010ab}.

Based on the success of AMANDA, the IceCube detector has been designed\cite{Ahrens:2003ix}.
In total $5160$ large area optical sensors have been deployed in the Antarctic ice at the geographical South Pole. They detect the Cherenkov light
produced by secondary leptons and hadrons as a result of charged current (CC)
and neutral current (NC) neutrino-nucleon interactions inside and outside the instrumented volume. The instrumented depth ranges from $1450$\,m to $2450$\,m  in the  ice. The sensors are attached to  86 vertical cable strings 
with $60$ sensors each and have a horizontal spacing of about $125$\,m between strings on a hexagonal grid.
 Along the strings the spacing is about $17$\,m, resulting
in  about  $1 \unit{km}^3$ of instrumented volume.
IceCube was completed  in its final configuration in December 2010 
and fully  commissioned
in May 2011. Already in its earlier configurations, based on the partly
installed detector, a substantial exposure was accumulated and first 
indications for an astrophysical signal were obtained\cite{Aartsen:2013vca}\cite{Aartsen:2013eka}.

\section{Summary of detection signatures}

The detection of neutrinos from cosmic accelerators, see contributions by 
Mészáros, K. Murase, E. Waxman, P. Lipari and M. Ahlers, 
is mainly based on their hard energy spectrum which is expected to 
follow the spectrum of accelerated primary cosmic rays and thus is 
expected to follow  a power-law with a hard spectral index
$\phi  \simeq \phi_0 \cdot E^{-2} $.
The largest signal is expected from close below the horizon and above, as the Earth becomes almost opaque to neutrinos 
above $\sim 100 \unit{TeV} -1 \unit{PeV}$.
Backgrounds are cosmic ray induced atmospheric muons and  neutrinos, which however exhibit at high energies a substantially softer spectrum.
Other powerful methods to detect astrophysical neutrinos above this background rely on  directional and time-correlations of the measured events. 
However, this requires strong individual sources. Alternatively, the cumulative flux of all cosmic sources is expected to  exceed the atmospheric 
backgrounds at high energies of typically $100 \unit{TeV} $.
 In this paper we focus on the detection of diffuse fluxes.
Depending on the luminosity density and strength of the sources, this approach
is very promising for the detection of a population of  
abundant but individually weak 
extra-galactic sources. Detailed discussions can be found e.g. in Ref.~\refcite{Lipari:2008zf}\cite{Kowalski:2014zda}\cite{Ahlers:2014ioa}.

Searching for diffuse neutrino fluxes at high energies requires a rigorous
 rejection of the overwhelming  atmospheric muon background and a 
precise modelling of the  partly irreducible  atmospheric
neutrino backgrounds.
The atmospheric muon background which penetrates from the surface to the depth of the detector can be rejected by focusing on up-going events
and/or events that interact within the instrumented volume.  
Atmospheric neutrinos can only be rejected if they arrive in the detector
from  above 
and the  corresponding air-shower is either tagged by  a 
surface detector\cite{Auffenberg:2014hqa} or by observing 
correlated atmospheric muons\cite{Schonert:2008is}.
For both types of background the signal to background ratio increases
with the energy threshold of the event selection.
%is a powerful strategy to enhance the 
%signal to background ratio and lower the rejection requirement.

Based on these signatures, neutrino telescopes can be sensitive to all
neutrino flavors, in particular when combining different detection 
channels and strategies. The basic signature of muon neutrino is a high-energy 
muons track from deep-inelastic CC neutrino-nucleon interactions. 
Electron neutrinos produce an electromagnetic cascade superimposed by a 
hadronic cascade at the interaction vertex. The lenght scale of the cascades is
small compared to the spacing of detector sensors. These events
are called cascade-like events. Tau neutrinos mostly produce a cascade signature very similar to electron neutrinos with two exceptions. First, at high energies above  a PeV the tau travels typically $50 \unit{m}\cdot E_\tau/\unit{PeV}$ 
before it decays. This results in a
characteristic  signature of two spatially separated cascades, called double-bang\cite{Learned:1994wg}. At all energies the tau may 
decay leptonically into a muon with a branching ratio of $\sim $\,17\%  
contributing to the track-like signature of muon neutrinos. 
All  flavors contribute equally to cascade-like events
via NC interactions.

\section{Observational status of different detection channels}

\subsection{High-energy starting events}

Remarkably, the discovery of an astrophysical neutrino signal by IceCube was achieved not in the muon channel that has been  the
intuitively assumed baseline channel
for decades %since Markov 
but with a new type analysis: the high-energy 
starting event analysis (HESE)\cite{Aartsen:2013jdh}\cite{Aartsen:2014gkd}.

The HESE analysis searches for neutrinos interacting inside 
the detector and is as such 
sensitive to all neutrino flavors and the full sky.
The selection splits the detector into an outer region, which is used
to tag and veto muon backgrounds from outside and an inner fiducial mass
of about $0.4 G\unit{t} $. Accepted events are required to deposit a visible
energy of more than %6000 photo electrons which is 
$20-30$ TeV inside the detector. Furthermore the earliest observed
 Cherenkov photons need to have been recorded within the fiducial volume
while  in the veto region no early 
signal in excess of the noise level is allowed.
The  analysis is based on the combination of essentially four innovative 
new methods that were not available during the operation of AMANDA and were 
not available during the design and construction phases of IceCube.
%These are:
\begin{romanlist}
\item It was realized that a search for
starting events would allow for a very simple all-sky search of all flavors, 
with a high significance because a large fraction of the energy is 
deposited at the interaction vertex. 
This has greatly improved the sensitivity, e.g. with respect to 
the single flavor up-going muon channel.
\item The ability to reasonably 
reconstruct the direction and energy of cascade-like events 
including a good estimate of the uncertainty based on the precise analysis of 
measured photomultiplier waveforms. This allowed quantifying the significance
of each event as the backgrounds strongly depend on the observed zenith angle.
\item  The usage of the outer layers of IceCube as veto allowed to model-independently quantify the remaining atmospheric muon background with good 
precision based on experimental data. It has  been shown that the background 
related to the inefficiency of the veto
falls off rapidly with energy and becomes insignificant 
above $\sim 60 \unit{TeV}$.

\item It was realized that the method of an atmospheric neutrino veto\cite{Schonert:2008is} could be successfully applied, greatly reducing the atmospheric 
neutrino background in the down-going region. The corresponding angular 
distribution is particularly important to unambiguously reject the 
hypothesis of a purely atmospheric origin, in particular as the high-energy atmospheric
neutrino flux from prompt decays of heavy quarks is largely uncertain\cite{Enberg:2008te}.
\end{romanlist}
It is the combination of all four methods in the interpretation of the observation, that made the detection of 
a cosmic neutrino signal evident and allowed to reject the hypothesis of atmospheric origin with high confidence.
As a side remark, this underlines the importance of a not too specific
optimization of large scale instruments which aim to explore unknown physics.
The implementation of potentially not fully optimized but 
 multi-purpose instruments which deliver higher data and information quality than minimally required allows a large flexibility in methods and fosters 
unforeseeable innovations which evolve only during the operation of the instrument.

\begin{figure}[ht]
\centerline{
\subfigure[Measured deposited energy]
{\includegraphics[height=2.in]{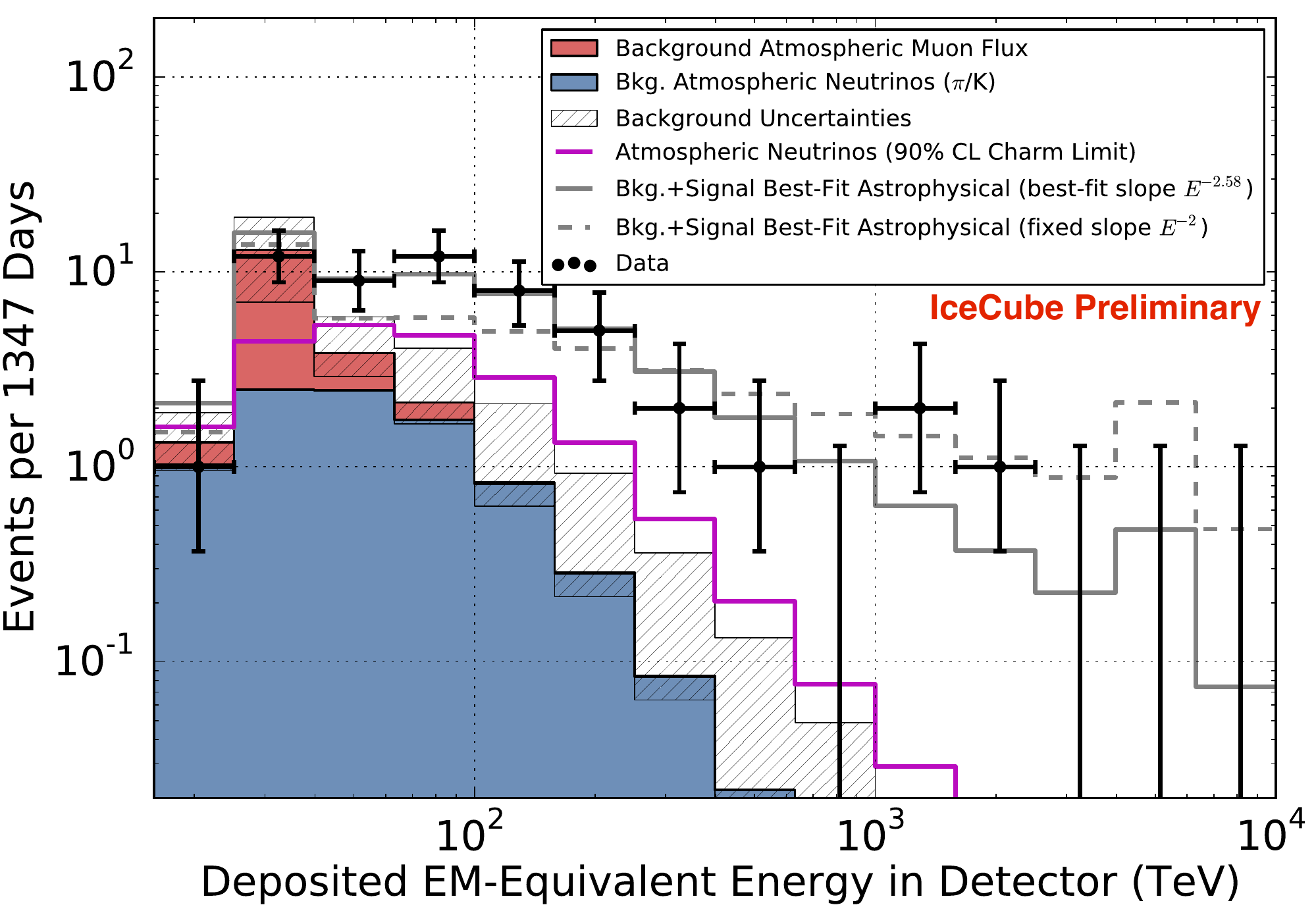}\label{fig:hese:dep}}
\hspace*{4pt}
\subfigure[Measured zenith]
{\includegraphics[height=2.in]{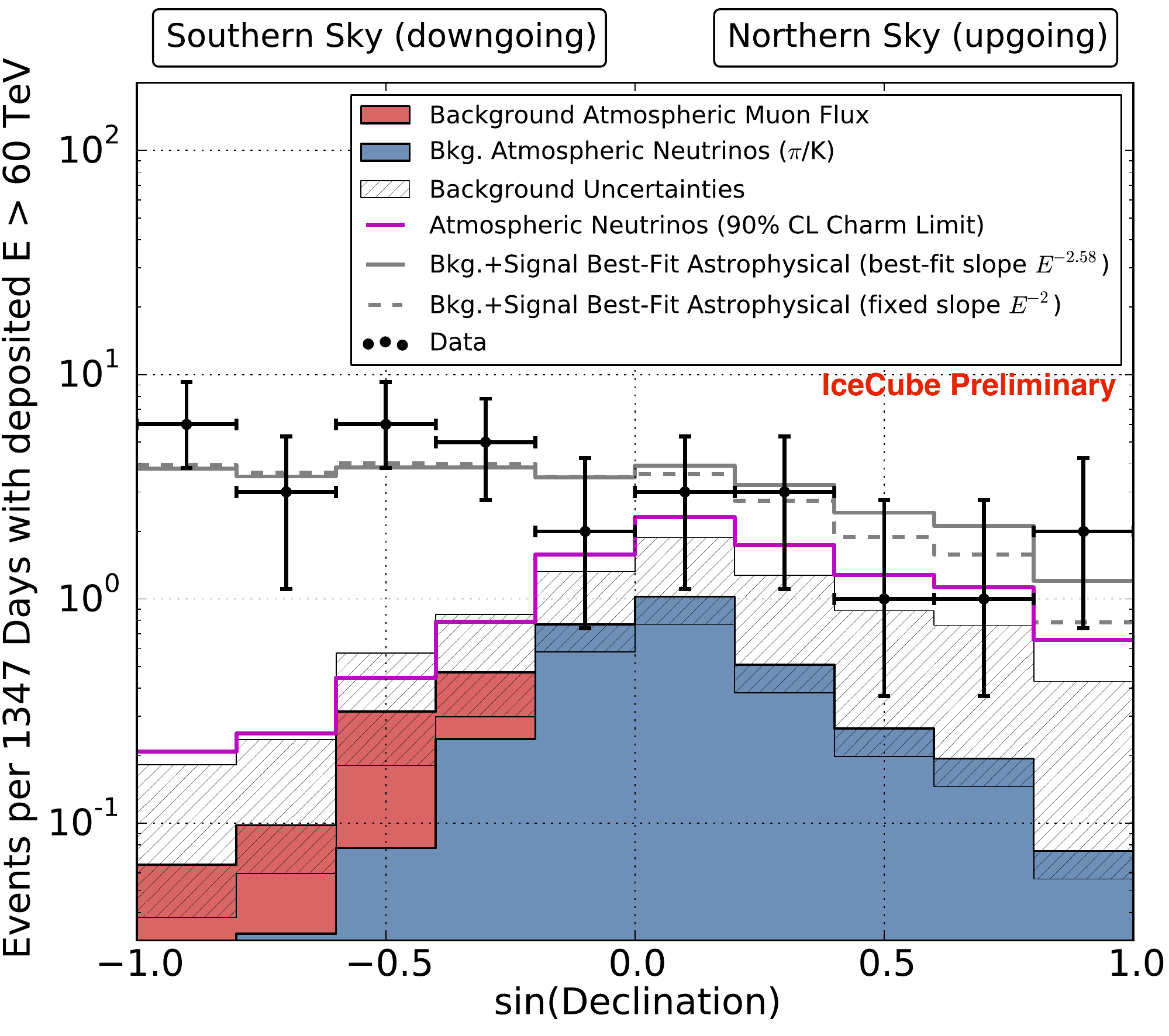}\label{fig:hese:fit}}
}
\caption{Results for the high-energy starting event analysis} \label{fig:hese4en} % common label
\end{figure}

Currently, data from three years of operation 
have been published\cite{Aartsen:2014gkd}. A pure atmospheric origin of the observed signal is rejected
with a confidence  level of $5.7 \sigma $.
Most recently, a fourth year of  data has been preliminarily released\cite{ICRC15HESE4} that
further increases the significance of the observation to $6.5 \sigma $. 
The observed distributions of energy and zenith, see fig. \ref{fig:hese4en}, agree well with
the expectation of a hard astrophysical spectrum and is clearly not compatible
with the expectation from atmospheric backgrounds.
The per flavor astrophysical flux measured with three years of data is
$ E^2 \phi(E) =0.84 \pm 0.3 \times 10^{-8} \unit{GeV cm^{-2} s^{-1} sr^{-1}} $, assuming a spectral index $\gamma=2 $ in the energy range between 
$60\unit{TeV} $ and $2 \unit{PeV}$.
A significant clustering of event directions has not been observed.

A particularly interesting extension of this analysis is to lower the 
energy threshold for this search. 
This is achieved by a gradually increasing
the veto thickness.
The loss in fiducial volume is compensated by larger fluxes at lower energy.
The first analysis of this type\cite{Aartsen:2014muf} using two years of data
has been able to lower the energy threshold to about  $1\, \unit{TeV} $ and extract
the astrophysical signal  down to about $10\, \unit{TeV} $.

\subsection{Cascade channel}

Closely related to the analysis of starting events are searches 
dedicated to cascade type events. Here, based on the event reconstruction, 
track-like event topologies are specifically rejected and a 
reasonably pure cascade-like sample is obtained. Backgrounds are
atmospheric muons with catastrophic %bremsstrahlung 
energy losses
that outshine the muon and thus  mimic a cascade-like signature. These  
can be suppressed by requiring containment of the interaction vertex
similar to the HESE analysis. The flux from conventional 
atmospheric electron neutrinos 
is considerably smaller (about a factor 20) than that of 
muon neutrinos and hence 
poses a relatively small background. The largest background 
uncertainty arises from 
prompt atmospheric neutrinos. Advantages of this analysis are a good
energy resolution for these contained cascades of the order of the energy scale uncertainty\cite{Aartsen:2013vja} $\simeq 10$\% and a lower energy threshold
 compared to the muon-channel. % which will be discussed below.
Cascade-analyses are usually sensitive to electron and tau neutrinos by 
CC interactions with a small contribution of NC interactions by all flavors.

Already a pioneering analysis, based on the configuration with only 40 installed detector strings found an excess of events above the atmospheric expectation\cite{Aartsen:2013vca}.
Recently, the first year data of the completed detector has been analyzed.
The energy spectrum of atmospheric electron neutrinos was measured to be consistent with the theoretical expectation.  No indication of a prompt signal was found and  the astrophysical component at high energies\cite{Aartsen:2015xup}
was confirmed.
The most recent results for the measurement of the astrophysical flux, based
on two years of data, are reported in Ref.~\refcite{ICRC15CASC}.
This analysis substantially increases the number of observed high-energy events
by also including partially contained events. 
It largely confirms the findings of the HESE analysis
and observes a cosmic signal with significance of more than $4\,\sigma$.

%AD09: High energy astrophysical neutrino flux characteristics for neutrino-induced cascades using IC79 and IC86-string IceCube configurations
%Author: Mariola, Achim, Hans
%We have performed a new measurement of the all-sky diffuse flux of high energy, E>10TeV, extraterrestrial neutrino induced showers (cascades) based on IceCube data collected during 641 days in 2010--2012. Cascades arise predominantly in electron and tau neutrino interactions and have good energy resolution, so that they are well-suited for the spectral characterization of the extraterrestrial flux. For the first time, we have included also high-energy cascades with vertices in near proximity to the detector, thereby enlarging the event sample by up to a factor of two for E > 100TeV. A total of 172 cascades with energies ranging from 10TeV to 1PeV have been observed, of which approximately 60% (75% above 100TeV) have not previously been reported by IceCube. Based on Monte Carlo simulations we estimate the neutrino purity to be 95%. The dominant extra-terrestrial component is well described by a smooth and featureless power-law. The result is in agreement with previous IceCube results and is preferred over a background-only hypothesis with a significance of more than 4 sigma. Additionally we will present a comparison between the results obtained when upward oriented and downward oriented showers are considered separately, showing that the extraterrestrial neutrino fluxes originating from the Northern and Southern hemispheres are consistent. 

\nopagebreak

\subsection{Muon channel\label{sec:mu_chan}}

The classical detection channel of neutrino telescopes is up-going
muon tracks  from CC  neutrino-nucleon interactions in and below the detector.
 The selection of up-going tracks efficiently eliminates the background of down-going cosmic ray induced atmospheric muons. The Earth's absorption  
increases
with energy and results in a zenith dependent expectation even for an isotropic 
diffuse cosmic signal. 
As the interaction can happen far outside the detector, the effective 
detection volume is much larger than the geometrical volume resulting in
larger  event rates than for contained events.
 However, muons from neutrino-interactions far away have lost a considerable and unknown fraction of 
their initial energy and carry only little information to 
distinguish astrophysical from atmospheric neutrinos.
In addition, %Another challenge for this channel is that
 through-going muons deposit only a small fraction of their energy
inside the detector. Therefore, the muon energy has to be estimated by the observed energy-loss, resulting in a resolution of about 
$\sim 50-70\% $ for the muon energy as compared to $ \sim 10 \% $ in case of contained cascades.
Diffuse searches using this channel contain potentially the largest number of cosmic neutrinos, but require larger data-sets to observe the same 
significance as channels with good energy resolution, e.g. contained cascades.
An important advantage of the muon channel compared to cascades is the 
good angular resolution, approaching about $0.1^\circ $ at high energies. 
Though angular information is not of primary concern 
in a diffuse search, it is 
helpful for an efficient rejection of the atmospheric muon 
background and the selection of high purity data-sets.

The analysis is done as a two-dimensional likelihood fit of the
measured energy  and  zenith angle. Fitting the full data 
 from a few $100$\,GeV  to high energies of a PeV
allows to strongly constrain systematic uncertainties.

The first indications of an astrophysical signal in the muon channels were found in the data of IceCube in the 59-string configuration\cite{Aartsen:2013eka}. Though the final significance of a cosmic signal was only $1.8 \sigma$ with respect to the conventional atmospheric background-only hypothesis, this observation has been important not only in the context of promising indications of a cosmic neutrino signal but also in its power to constrain the conventional and prompt atmospheric neutrino backgrounds for the analysis of starting events.
Particularly spectacular has been the highest energy muon, which energy in the detector has been estimated to 400\,TeV

Since then, the evidence in this channel has been steadily increasing.
A combined analysis using $35,000$ events from the 79 string and first year of 
86 string configurations (2010-2012) has found evidence for an astrophysical
flux above  $300$\,TeV
consistent with the HESE result  at the $3.7\,\sigma$ level\cite{Aartsen:2015rwa}.
Most recently, the full data from six years of IceCube operation including
the 59-string and 79 string configurations as well as four years
of IceCube with 86 strings (2009-2015) has been analyzed\cite{ICRC15LEIF}. 
Here, the event selection efficiency has been optimized resulting in a total 
of about $340,000 $ muon neutrino events with
an estimated  purity of better than $99.9\% $. 
The significance of an astrophysical flux with the full six years is
%and rejects the atmospheric- only hypothesis 
 at the level of $6 \sigma $ (rejecting a pure atmospheric origin).

\begin{figure}
\centerline{\includegraphics[width=6.5cm]{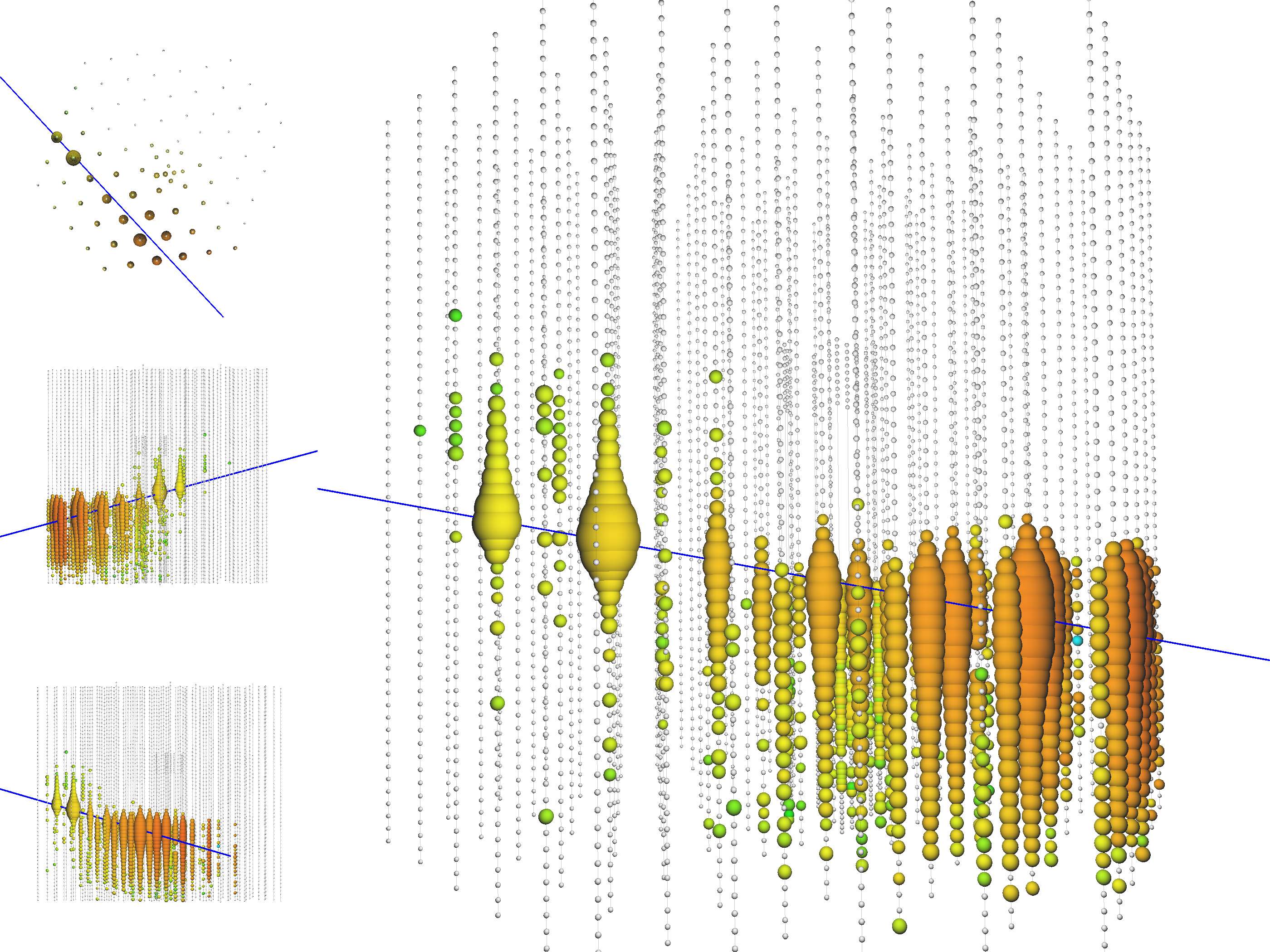}}
\caption{Event-view of the multi-PeV upgoing muon event detected with IceCube. Colored spheres indicated optical sensors that registered a signal, where the size encodes the logarithm of detected charge and the color the arrival time (red early and green late). Left are the three projected views. The reconstructed track is indicated as a line.}
\label{fig:kloppo}
\end{figure}

Also notable is the observation of an ultra-high-energy neutrino event\cite{schoenen2015detection},
%that we do not name \emph{Kloppo}, 
see Fig. \ref{fig:kloppo}.
It is a through-going track that deposits an energy of $2.6 \pm 0.3 \,\unit{PeV} $ within the instrumented volume of IceCube. Based on simulations of 
events with similar topology such an energy loss would be expected from a muon of more than $4 \,\unit{PeV}$ and thus implies an even higher neutrino energy. This makes this event the highest-energy neutrino detected %by IceCube 
to date. Based on the huge
energy, for this event alone the hypothesis of an atmospheric origin can be rejected\cite{ICRC15LEIF}
for this event alone by about   $4 \sigma$. Though not relevant for 
the subject of this report, it is worthwhile to note that the directional 
uncertainty is less than $0.3^\circ $, but attempts to identify an astrophysical source have not been successful yet. The closest known source of GeV
  photon emission\cite{Acero:2015gva} is about $3^\circ $ away and 
$11^\circ $ for known TeV sources\cite{Wakely:2007qpa}. The direction is $11^\circ $ off the Galactic Plane.

\subsection{EHE channel}

In this channel neutrinos of extremely high energies are searched for.
This channel targets is very large energy depositions related to
events of typically $10^{17} \unit{eV} $ energy,  
as e.g.\ expected from the GZK effect.
Because of the decreasing background of cosmic ray induced atmospheric 
muons the energy threshold can be gradually reduced towards the horizon.
For straight down-going events, the surface detector IceTop is added as a veto.
As a consequence of absorption within Earth  the region around 
and above the horizon is  particularly important.

It was this type of analysis that initially observed the first neutrino 
events with 
PeV energy\cite{Aartsen:2013bka} close to 
its energy threshold. 
The analysis has been based on an exposure of one year in the 79 string configuration and  the first year of full IceCube operation. An update to this analysis\cite{AYAICRC} to 6 years of IceCube operation has further increased 
 the sensitivity, particularly towards high energy.
No further events with energies substantially above a few PeV have been observed which results in the currently most constraining exclusion limits for ultra-high-energy neutrinos such as expected from the GZK effect\cite{Beresinsky:1969qj}.
As no further events of higher energy have been observed with recent data 
 and thus no improved spectral information on the diffuse flux 
can be deduced, this channel is ignored in the following discussions.

\subsection{Tau channel}

The tau neutrino is interesting because due to oscillations about one third of astrophysical neutrinos are expected to be of tau flavor. Furthermore, due to ``regeneration'' in tau decays, the Earth is not fully opaque to tau neutrinos\cite{Beacom:2001xn}.
The atmospheric background of high energy tau neutrinos 
is substantially smaller than even that of prompt neutrinos\cite{Gondolo:1995fq},  and thus any observed a tau neutrino would be 
astrophysical with high probability.

As discussed above, the detection signature of tau neutrino interactions is
mostly similar to electron  neutrinos unless their energy exceeds about a 
PeV. Nonetheless it is interesting to attempt the identification of a double-bang signature in the sample of 
observed starting events. This has been performed with a modification\cite{hallenmeasurement} of the 
reconstruction algorithm used for starting events\cite{Aartsen:2013vja}.
No evidence of a double-bang 
signature has been found. However, this analysis, close
to the detection threshold, is challenging because of 
 systematic uncertainties of the photon propagation through ice and a substantial contribution of tau neutrinos to the observed events cannot be excluded.
A more robust approach is to directly search for large double pulse signatures in the recorded waveforms of optical sensors. An independent dedicated
 search\cite{ICRC15TAU} for this signature has not detected a clear double-bang event. However, the sensitivity of this search has not yet reached the observed astrophysical flux level and more data is needed.

\section{Comparison of observational results}

\subsection{Energy spectrum}

\begin{figure}[ht]
\centerline{
\subfigure[Measured deposited energy]
{\includegraphics[width=2.7in]{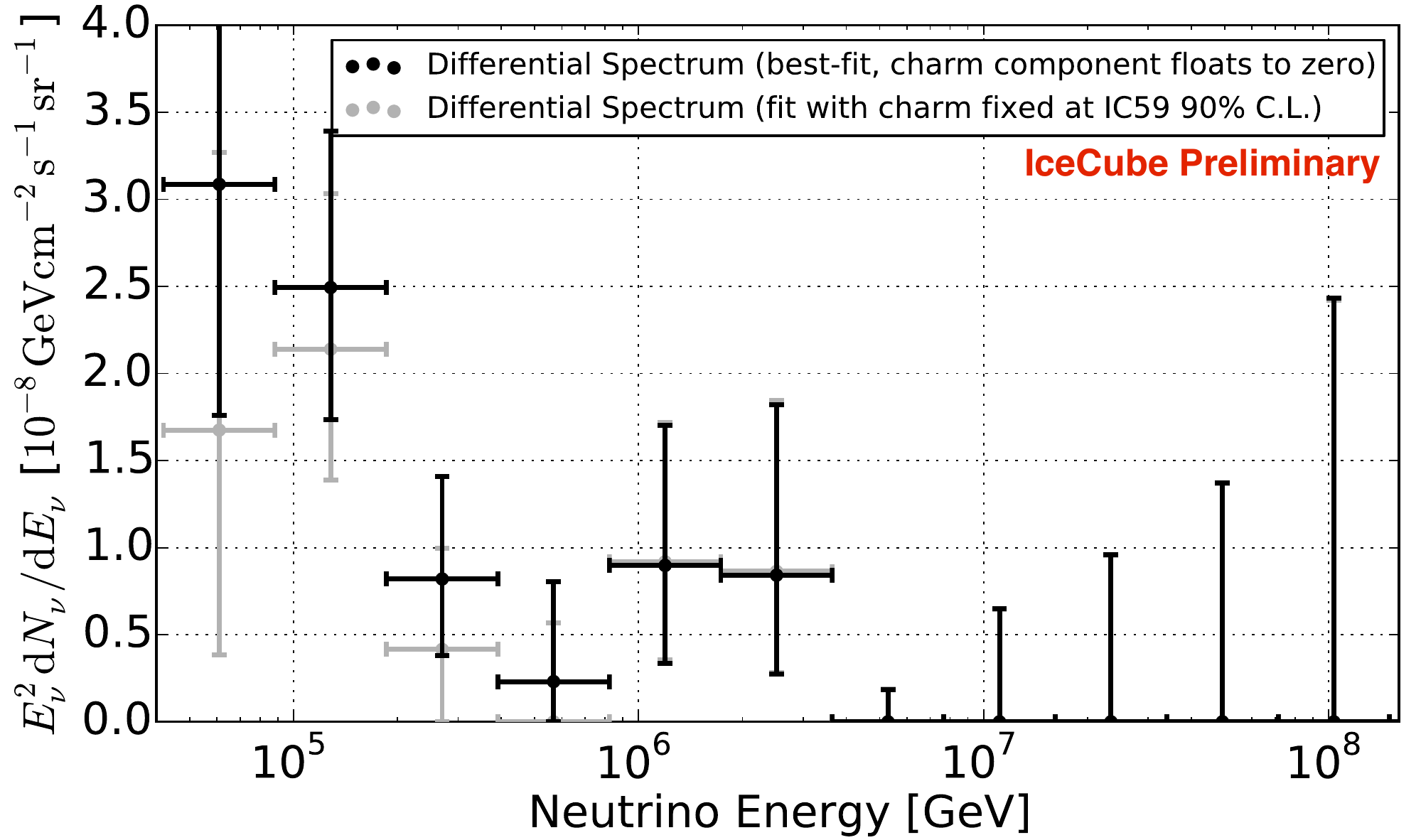}\label{fig:hese:spec}}
\hspace*{4pt}
\subfigure[Reconstructed energy spectrum]
{\includegraphics[width=2.2in]{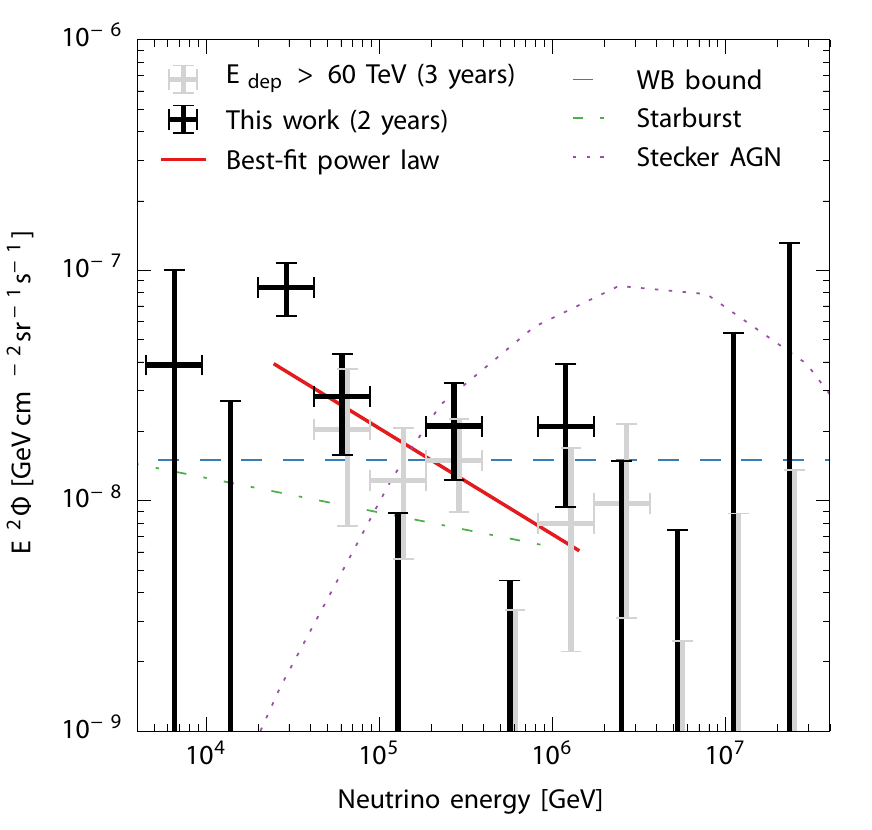}\label{fig:jacob:fit}}
}
\caption{Results for the high-energy starting event analysis} \label{fig:hese4enrec} % common label
\end{figure}

The  energy spectrum measured with the most recent data\cite{ICRC15HESE4}
for high-energy starting events is shown in Fig. \ref{fig:hese4enrec} together
with the spectrum obtained from the  extension of the analysis towards 
lower energies\cite{Aartsen:2014muf}.
Here, the normalization in each energy bin has been a free parameter
in a maximum likelihood fit of the data-set. 
The best fit spectral index for Fig. \ref{fig:hese:spec} is $\gamma =2.58\pm 0.25 $, slightly softer than results based on earlier data.
This is consistent with the result in Fig. \ref{fig:jacob:fit} of 
$\gamma =2.46\pm 0.12 $ and is also consistent with measurements in the cascade channel\cite{ICRC15CASC} which find similar soft indices. 
 The spectrum slightly depends on the assumptions of atmospheric neutrinos from charm decays. 
A larger charm component would lead to a 
harder astrophysical spectrum. 

Two aspects are particularly interesting. On may wonder about the existence
of  a possible cut-off
at an energy of a few PeV or  a possible spectral break in the spectrum.
Note, that the under-fluctuation of events just below a PeV deposited energy 
has previously caused some speculations. However, this fluctuation
 is not statistically significant and it has decreased 
with the recently added data.

Clearly visible is the deviation from the hard $E^{-2} $ hypothesis.
It seems that the steepness of the slope is dominated by data at 
lower energy 
$\lesssim 50 \unit{TeV} $. As shown in Ref.~\refcite{Aartsen:2014muf}, above an energy of $100 \unit{TeV} $ the data would be well consistent with also a hard spectrum $\gamma \simeq 2.26\pm 0.35 $.

For the question of a cut-off 
 about $3$ events would be expected for a hard $E^{-2} $ spectrum above 
$2$\,PeV, while none were observed.
However, with a softer spectrum, as the best fit seems to indicate, 
this tension is strongly relaxed and a 
cut-off is not required to describe the observation.

In conclusion, the current statistics is not sufficient to answer
  questions concerning a possible spectral break
or  cut-off and future data  will improve the picture.

\begin{figure}[ht]
\centerline{
\subfigure[Unfolded neutrino energy distribution]
{
\includegraphics[width=2.7in]{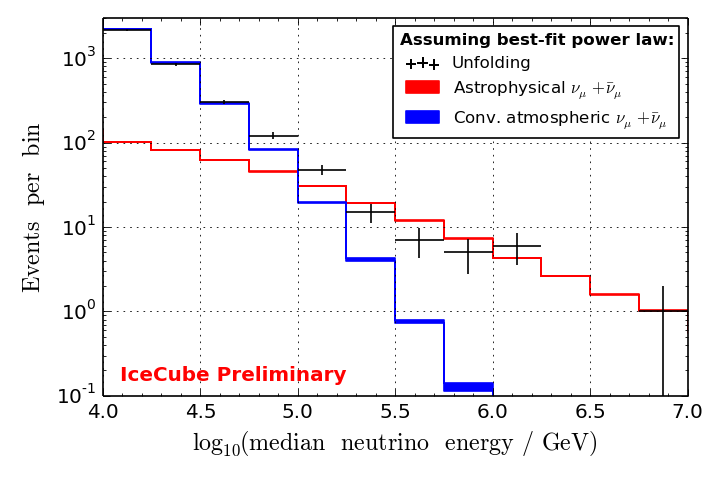}
\label{fig:leifenergy}}
\hspace*{4pt}
\subfigure[2-d profile likelihood of the fit spectral index versus flux]
{
\includegraphics[width=2.3in]{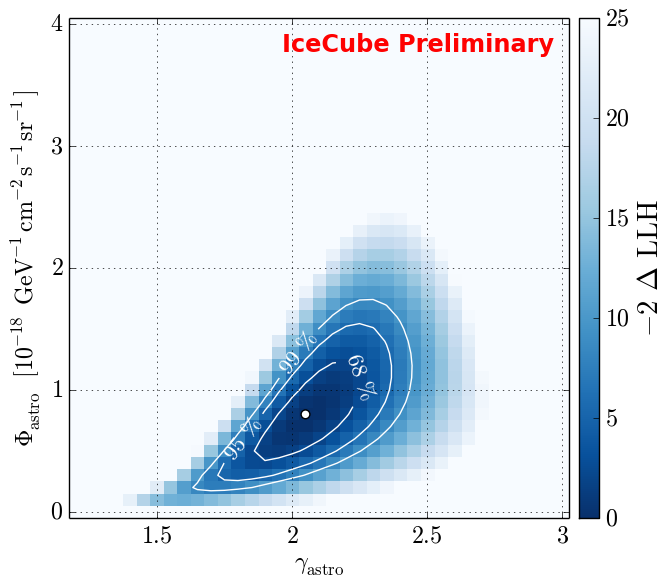}
\label{fig:leifindex}}
}
\caption{Results of the analysis of up-going muons based on six years data} \label{fig:upmu} % common label
\end{figure}

While the starting event analysis is dominated by cascade-like events from the Southern Sky it 
is interesting to compare to the energy distribution of up-going muon neutrinos.
As the measurement of uncontained muons does not directly allow for a good estimate of the neutrino energy, the spectral unfolding is difficult but benefits from higher statistics of events and well controlled systematics and 
backgrounds.
The measurement with two years of data\cite{Aartsen:2015rwa} resulted
in a spectral index  slightly harder $\gamma = 2.2 \pm 0.2 $ 
than measured in the cascade dominated channels. 
This tension has increased with 
 recent data covering 6 years of IceCube\cite{ICRC15LEIF}
as shown in  Fig.\ref{fig:upmu}. The left figure shows the 
reconstructed energy distribution, the right figure the profile likelihood for the extracted astrophysical flux parameters.
Note, that the reconstructed neutrino
energy spectrum is model dependent. Shown here is the median expected neutrino 
energy calculated from the measured deposited energy % each event 
assuming the best fit spectrum.
The excess of data is clearly inconsistent with a pure atmospheric origin.
With a best fit spectral index of $2.06\pm 0.13$ 
the observed spectrum is harder than that of the cascade dominated measurements.

For a quantitative discussion the results for all analyses have to be compared
by means of statistical confidence. % as shown in Fig.\ref{fig:leifindex}. 
It is found that the 
measured flux normalization is often correlated with the fitted spectral
index. When analyzing the error
contours for all analyses\cite{HANSSEBINT} the apparent tension is reduced
to about $2.5 \sigma $ which is marginally statistically significant.
For this discussion  it is furthermore important to highlight the systematic differences between these two measurements. The threshold for the up-going muon signal
 is a few $100\unit{TeV} $ while 
astrophysical starting events are detected above a few $10 \unit{TeV} $.
If only high-energy starting events were considered in the comparison,
 the spectra would be in
agreement. Another important difference is the dominance of different 
hemispheres in both analyses. If the astrophysical flux was non-isotropic or e.g.\ composed of a 
galactic and an extra-galactic component a difference between the two analyses
could be explained.
Additional data will be needed to answer these questions. An extension to this 
discussion is found in Sec.\ref{sec:isotropy}.

\subsection{Global fit and flavor ratio}

\begin{figure}[ht]
\centerline{
\subfigure[Energy spectrum]
{\includegraphics[width=2.5in]{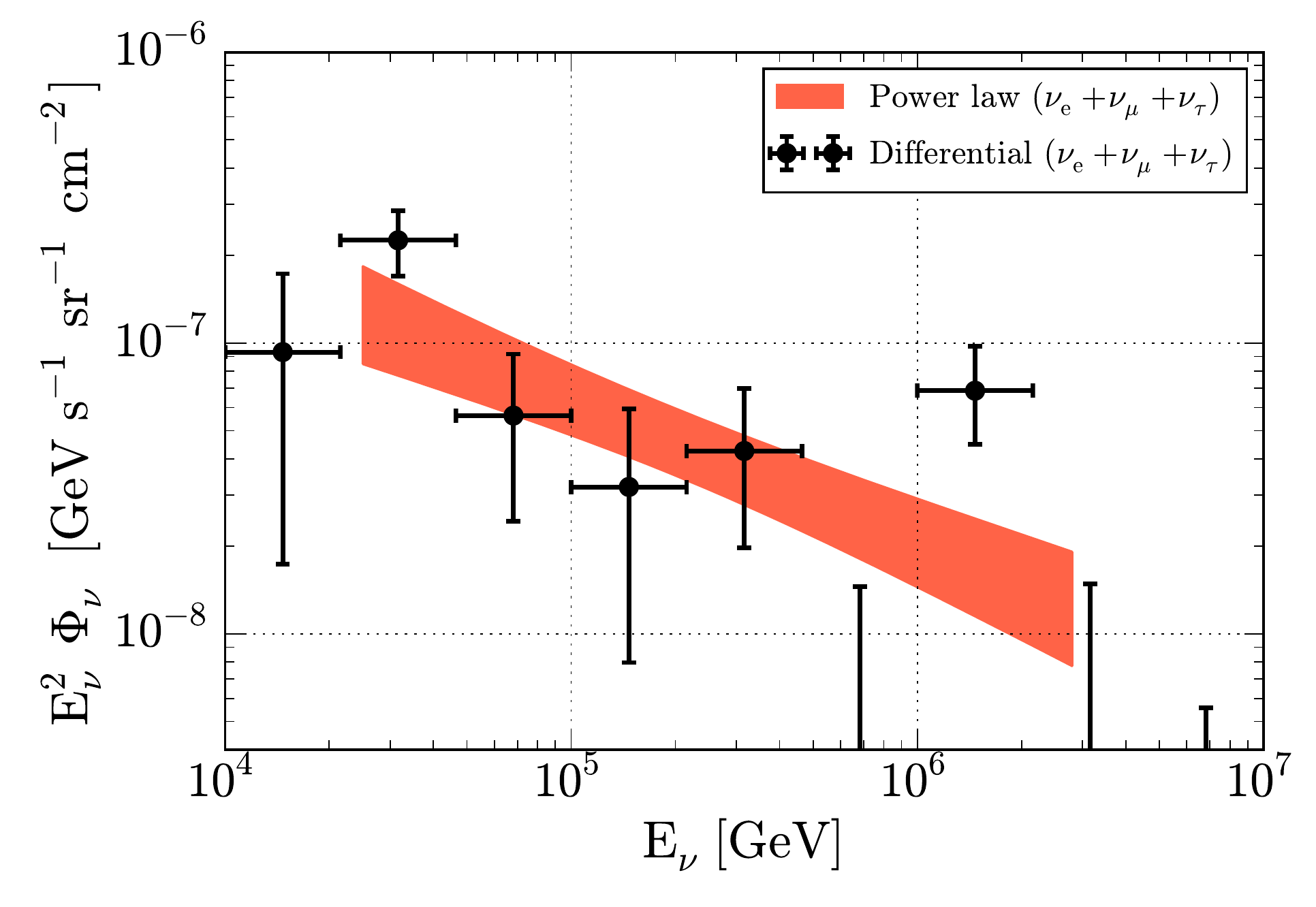}\label{fig:global:spec}}
\hspace*{4pt}
\subfigure[Flavor triangle]
{\includegraphics[width=2.5in]{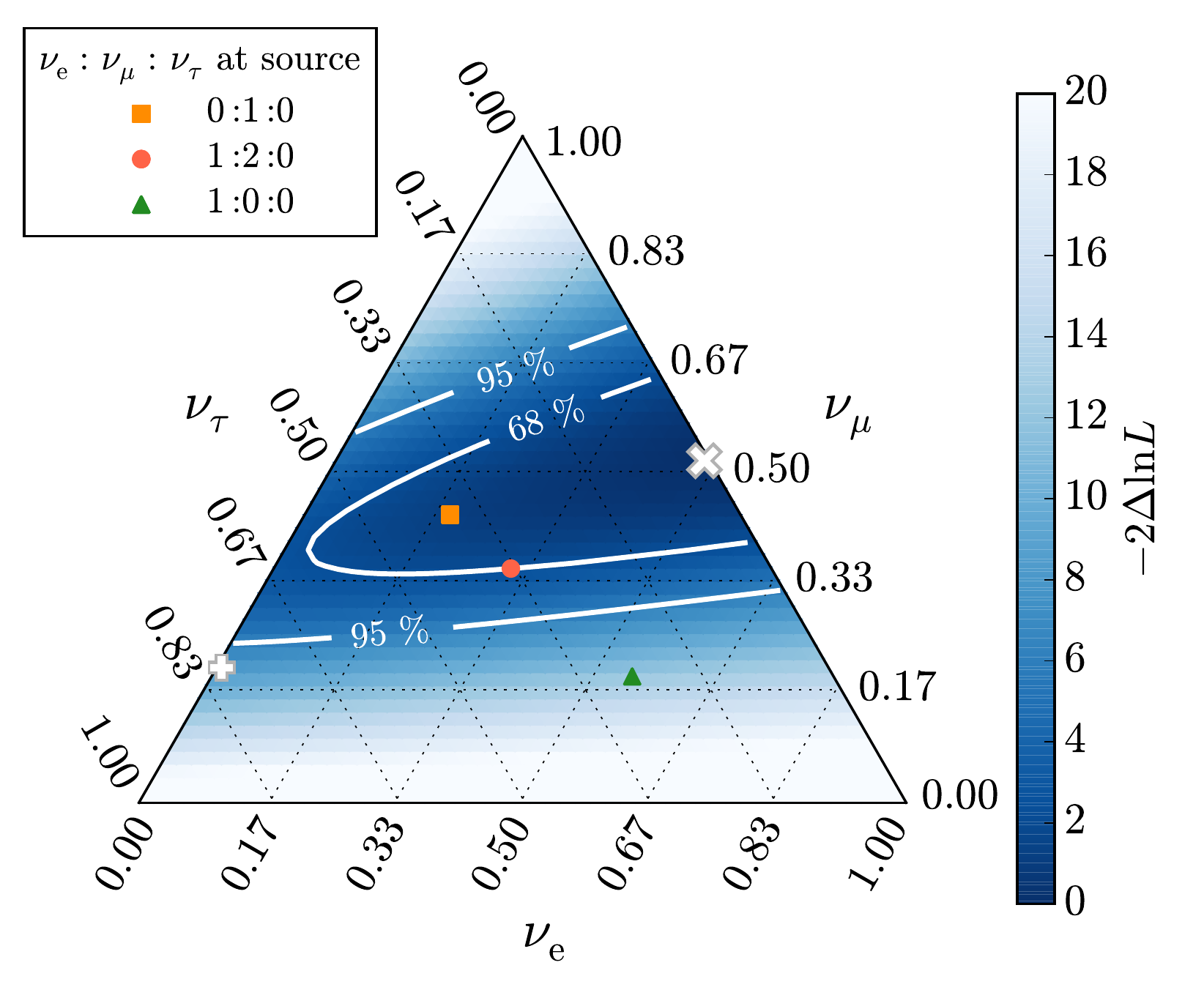}\label{fig:global:flavor}}
}
\caption{Results of the global fit} \label{fig:global} % common label
\end{figure}

Based on the different detection channels a global fit of the combined data-sets can be attempted.
In Ref.~\refcite{Aartsen:2015knd} such a global analysis has been performed using six different data-sets\cite{Aartsen:2013eka}\cite{Aartsen:2015rwa}\cite{Aartsen:2013vca}\cite{Aartsen:2013jla}\cite{Aartsen:2013jdh}\cite{Aartsen:2014gkd}\cite{Aartsen:2014muf} consisting of up-going muons, contained cascades 
and starting events.
Special care was taken not to double count overlapping data-sets and the combination
of  systematic uncertainties. For practical reasons, 
the fit does not include the full systematics of the individual data-sets but combines them as generalized global parameters. These are the energy scale uncertainty, the atmospheric muon background normalization for each data-set and
the cosmic ray spectral index which affects the atmospheric neutrino background. Different hypotheses for the astrophysical flux normalization and spectra are tested.
The result for a single power-law is  a spectral index of $\gamma = -2.50 \pm 0.09 $, disfavoring a hard spectral index of 
$\gamma =2$ at the $3.8\sigma $ level. This significance is reduced to $2.1 \,\sigma $ if an exponential cut-off is introduced.
Similar to the discussions above, it can be seen in  Fig. \ref{fig:global:spec}, that this tension could be also relaxed
without introducing a cut-off  if a transition from a softer spectral 
index  at lower energies to a harder spectrum at higher energies is assumed.

The flavor ratio of the neutrino flux is particularly interesting, because 
it allows to constrain the acceleration mechanism. The initial 
flavor composition at the source is modified by neutrino oscillations and
largely smeared out due to the long baseline. Therefore, one expects to
observe all neutrino flavors at Earth with a similar flux.
However, depending on the injection model of neutrino flavors at the source
and the assumed oscillation parameters deviations from
the exact $\nu_e:\nu_\mu:\nu_\tau \approx 1:1:1 $ mixing are expected at Earth.
Despite of not having directly identified tau neutrino %$\nu_\tau $ 
events, they contribute to
the observed event rates of the considered detection channels differently.
Within the framework of the global fit the flavor ratio can be tested when fitting
the flux normalization of each flavor seperately in a joint fit. 
The result is shown in Fig. 
\ref{fig:global}. The measured data is consistent with a mixture
of all flavors and pure fluxes of $\nu_e $ and $\nu_\mu $ or $\nu_\tau $ are strongly disfavored. When comparing to different injection scenarios
the hypothesis of 
pure $\nu_e $ as expected from sources with dominating muon decays can be excluded. Both scenarios of an injected ratio $\nu_e:\nu_\mu:\nu_\tau \approx 0:1:0 $
and $\nu_e:\nu_\mu:\nu_\tau \approx 1:2:0 $ are compatible with the data.
With more data  this measurement is expected to become increasingly important for the understanding of the source mechanism of the measured flux.

\subsection{Isotropy\label{sec:isotropy}}

No analysis of the arrival directions of detected neutrinos
has  yet revealed a statistically significant clustering nor correlation with a known source\cite{ICRC15HESE4}\cite{Aartsen:2014cva}\cite{Aartsen:2014ivk}. However, also tests for large-scale anisotropies allow to investigate the important  question
whether the  assumption of  isotropy of the cosmic signal is valid or
if regions related to a few close sources dominate.
In particular, the question whether the observed flux is 
of galactic or extra-galactic origins 
can be tested. 

An important step is the recent observation of the astrophysical signal 
also in the up-going muon analyses\cite{Aartsen:2015rwa}\cite{ICRC15LEIF}, see  Sec.\ref{sec:mu_chan}. This confirms that the flux that was observed in the high-energy starting event analysis mostly by cascades from the Southern hemisphere is  accompanied by a roughly equally strong flux of muon neutrinos in the 
Northern hemisphere. This indicates
 that at least a substantial fraction of the flux is isotropic and thus presumably extra-galactic.

\begin{figure}
\centerline{
\includegraphics[width=4.7cm]{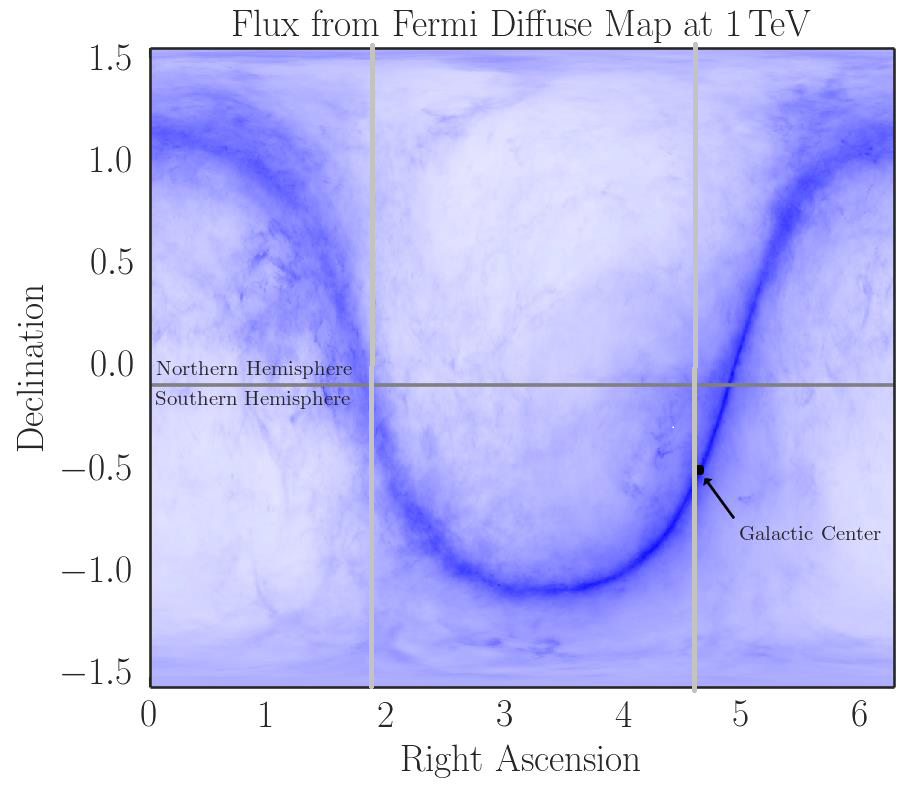}}
\caption{Splitting the data samples into regions of the sky: North-South and East-West. The figure shows the orientation of the galactic plane, indicated by the superimposed diffuse gamma emission measured by Fermi-LAT, \texttt{http://fermi.gsfc.nasa.gov/ssc/data/access/lat/BackgroundModels.html}. %, \texttt{gll\_iem\_v05\_rev1.fit}). 
Vertical lines indicate a split by right ascension that results in quadrants
in both hemispheres with and without the galactic plane.
}
\label{fig:fermi}
\end{figure}

\begin{figure}
\centerline{
\includegraphics[width=9.9cm]{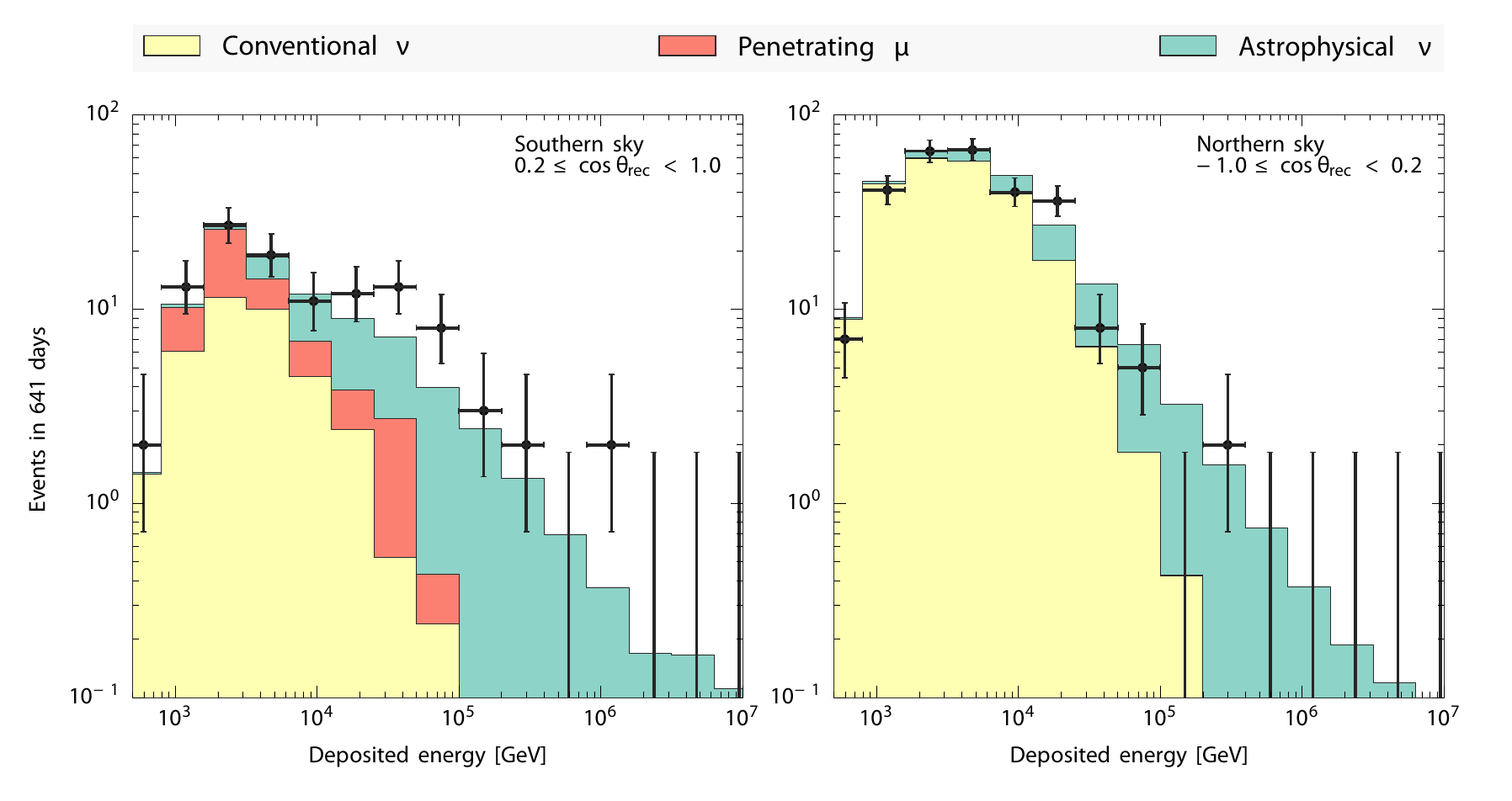}
}
\caption{Energy spectra of starting events for Northern and Southern Sky separately.}\label{fig:north_south}

\end{figure}

A straight forward approach is  to split the data samples into two separate regions of the sky and to compare the observed fluxes.
This is indicated in Fig.\ref{fig:fermi}.

The simplest test is to compare the Northern and Southern hemispheres. As the
Southern hemisphere contains the central part of the Galaxy, 
differences in the hemispheres may allow to indicate differences  between 
galactic and extra-galactic components. 
This has been done within
the analysis of starting events at lower energy threshold\cite{Aartsen:2014muf}, see Fig.\ref{fig:north_south}, the global fit\cite{Aartsen:2015knd} and the analysis of contained cascades\cite{ICRC15CASC}.
Both hemispheres show a clear excess of an astrophysical signal 
over the atmospheric background, however, the southern excess seems stronger.

 This question  is more quantitatively addressed in the aforementioned global fit.  It results 
in a spectral index of $\gamma = 2.0_{-0.4}^{+0.3} $ for the Northern Sky 
and  $\gamma = 2.56\pm 0.12 $ for the Southern Sky, respectively.
Note, that the significance of a discrepancy is only $1.1 \,\sigma$.
The interpretation has to be done carefully, as the observational conditions
and systematic uncertainties between north and south strongly differ.
The fit of the Northern Sky is dominated by the up-going muon sample which has a higher energy threshold compared to the cascade sample, which dominates the Southern Sky. Furthermore, detector systematics with all sensors 
directed down-ward, 
 the absorption of high-energy events by Earth, the rejection of 
atmospheric neutrino background differs strongly between both hemispheres.

This test has been repeated with contained cascades\cite{ICRC15CASC}.
This measurement results in very similar spectral indices for North ($\gamma = 2.69_{-0.34}^{+0.34} $) and South ($\gamma = 2.68_{-0.22}^{+0.20}$). 
Also for this analysis systematics differ between North and South and the energy range is slightly smaller. Obviously, at this point in time the results are not conclusive yet.

\begin{figure}
\centerline{
\subfigure[Flux normalization]{
\includegraphics[width=5.8cm]{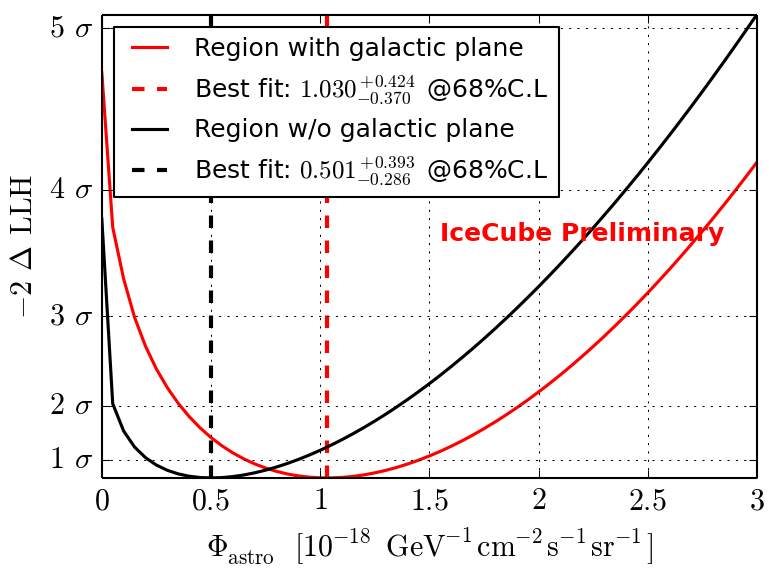}
}
\hspace*{4pt}
\subfigure[Spectral index]{
\includegraphics[width=5.3cm]{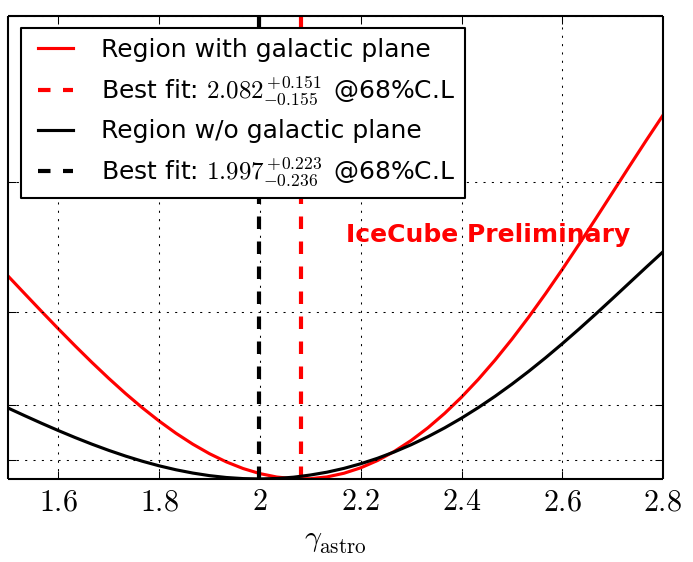} 
}
}
\caption{Profile likelihoods for the extracted flux parameters from 
the fit of the six year up-going 
muon sample which is split by right ascension.}\label{fig:rasplit}
\end{figure}

A test that is not affected by these systematics has been performed 
in the Northern Sky with the six years up-going muon sample. The sample was
split in right ascension instead of declination. The regions, indicated as vertical lines 
in Fig.\ref{fig:fermi}, are chosen such, that the two split samples
are of similar statistics but complementary with respect to the galactic plane.
The fit result, shown in Fig.\ref{fig:rasplit}, is a small but not statistically significant larger flux and softer spectrum from the region including  the galactic plane.

Again, a definitive answer whether the flux from the galactic plane differs from the all-sky result can not be deduced. 
Dedicated tests are underway and additional data will 
certainly allow to address this question further.

\section{Summary and conclusions}

In this contribution we have reviewed the observational evidence and approaches to observe diffuse fluxes of astrophysical neutrinos. It is remarkable that only a few years after the initial discovery, the signal could be clearly 
identified in several detection channels. 
After the consolidation of the observational evidence
substantial progress has 
been made in characterizing the signal properties, in particular by
improving analysis methods and by the comparison of the different results.
 Several important questions have been
addressed, especially the 
characterization of the energy spectrum, the flavor composition and 
a possible large-scale isotropy.
While we are still awaiting the hopefully soon detection of point sources (see contribution by Ch. Finley), 
the above questions   can be addressed with further improved analyses and new data.

Ultimately IceCube will be limited in sensitivity. Therefore, the need for 
a next-generation instrument arises. 
The KM3Net\cite{Margiotta:2014gza} neutrino telescope in the Mediterranean Sea (see  contribution by M. de Jong)
 and the BAIKAL-GVD neutrino telescope\cite{Avrorin:2013sla} in lake Baikal (see  contribution by Z. Djilkibaev)
plan to install deep
water instrumentations that exceed the size of the  IceCube detector. 
Finally, the design of a next generation instrument at the South Pole, IceCube-Gen2\cite{Aartsen:2014njl}, has started
aiming for the scale of $10 \unit{km}^3 $ volume (see contribution
by O. Botner).
Also investigations of possible improvements of  the current IceCube 
 performance  for the search of cosmic neutrinos by extending the surface detector IceTop acting as a veto\cite{Auffenberg:2014hqa} to atmospheric signals are underway.

\section*{Acknowledgements}
The author would like to thank Hans Niederhausen, Leif R\"adel and Sebastian Schoenen for careful reading of the manuscript as well as the whole 
IceCube collaboration for their contributions.

\bibliographystyle{ws-rv-van}
\bibliography{references}

\begin{thebibliography}{48}
\providecommand{\natexlab}[1]{#1}
\providecommand{\url}[1]{\texttt{#1}}
\expandafter\ifx\csname urlstyle\endcsname\relax
  \providecommand{\doi}[1]{doi: #1}\else
  \providecommand{\doi}{doi: \begingroup \urlstyle{rm}\Url}\fi

\bibitem{Markov:1960vja}
M.~Markov, {On high energy neutrino physics}, \emph{{Proceedings, 10th
  International Conference on High-Energy Physics (ICHEP 60)}}. pp. 578--581,
  (1960).

\bibitem{Achar:1965ova}
C.~V. Achar, M.~G.~K. Menon, V.~S. Narasimham, P.~V.~R. Murthy, B.~V.
  Sreekantan, et~al., {Detection of muons produced by cosmic ray neutrinos deep
  underground}, \emph{Phys.Lett.} {\bf 18}, \penalty0 196--199,  (1965).
\newblock \doi{10.1016/0031-9163(65)90712-2}.

\bibitem{Reines:1965qk}
F.~Reines, M.~Crouch, T.~Jenkins, W.~Kropp, H.~Gurr, et~al., {Evidence for
  high-energy cosmic ray neutrino interactions}, \emph{Phys.Rev.Lett.} {\bf
  15}, \penalty0 429--433,  (1965).
\newblock \doi{10.1103/PhysRevLett.15.429}.

\bibitem{Gaisser:1994yf}
T.~K. Gaisser, F.~Halzen, and T.~Stanev, {Particle astrophysics with
  high-energy neutrinos}, \emph{Phys.Rept.} {\bf 258}, \penalty0 173--236,
  (1995).
\newblock \doi{10.1016/0370-1573(95)00003-Y}.

\bibitem{Belolaptikov:1997ry}
I.~Belolaptikov et~al., {The Baikal underwater neutrino telescope: Design,
  performance and first results}, \emph{Astropart.Phys.} {\bf 7}, \penalty0
  263--282,  (1997).
\newblock \doi{10.1016/S0927-6505(97)00022-4}.

\bibitem{Balkanov:1997gw}
V.~Balkanov et~al., {Reconstruction of atmospheric neutrinos with the Baikal
  Neutrino Telescope NT-96}.  (1997).

\bibitem{Andres:1999hm}
E.~Andres, P.~Askebjer, S.~Barwick, R.~Bay, L.~Bergstrom, et~al., {The AMANDA
  neutrino telescope: Principle of operation and first results},
  \emph{Astropart.Phys.} {\bf 13}, \penalty0 1--20,  (2000).
\newblock \doi{10.1016/S0927-6505(99)00092-4}.

\bibitem{Andres:2001ty}
E.~Andres, P.~Askebjer, X.~Bai, G.~Barouch, S.~Barwick, et~al., {Observation of
  high-energy neutrinos using Cherenkov detectors embedded deep in Antarctic
  ice}, \emph{Nature}. {\bf 410}, \penalty0 441--443,  (2001).
\newblock \doi{10.1038/35068509}.

\bibitem{Achterberg:2007qp}
A.~Achterberg et~al., {Multi-year search for a diffuse flux of muon neutrinos
  with AMANDA-II}, \emph{Phys.Rev.} {\bf D76}, \penalty0 042008,  (2007).
\newblock \doi{10.1103/PhysRevD.76.042008, 10.1103/PhysRevD.77.089904}.

\bibitem{antares:2011nsa}
M.~Ageron et~al., {ANTARES: the first undersea neutrino telescope},
  \emph{Nucl.Instrum.Meth.} {\bf A656}, \penalty0 11--38,  (2011).
\newblock \doi{10.1016/j.nima.2011.06.103}.

\bibitem{Aguilar:2010ab}
J.~Aguilar et~al., {Search for a diffuse flux of high-energy $\nu_\mu$ with the
  ANTARES neutrino telescope}, \emph{Phys.Lett.} {\bf B696}, \penalty0 16--22,
  (2011).
\newblock \doi{10.1016/j.physletb.2010.11.070}.

\bibitem{Ahrens:2003ix}
J.~Ahrens et~al., {Sensitivity of the IceCube detector to astrophysical sources
  of high energy muon neutrinos}, \emph{Astropart.Phys.} {\bf 20}, \penalty0
  507--532,  (2004).
\newblock \doi{10.1016/j.astropartphys.2003.09.003}.

\bibitem{Aartsen:2013vca}
M.~Aartsen et~al., {Search for neutrino-induced particle showers with
  IceCube-40}, \emph{Phys.Rev.} {\bf D89}\penalty0 (10), \penalty0 102001,
  (2014).
\newblock \doi{10.1103/PhysRevD.89.102001}.

\bibitem{Aartsen:2013eka}
M.~Aartsen et~al., {Search for a diffuse flux of astrophysical muon neutrinos
  with the IceCube 59-string configuration}, \emph{Phys.Rev.} {\bf
  D89}\penalty0 (6), \penalty0 062007,  (2014).
\newblock \doi{10.1103/PhysRevD.89.062007}.

\bibitem{Lipari:2008zf}
P.~Lipari, {Proton and Neutrino Extragalactic Astronomy}, \emph{Phys.Rev.} {\bf
  D78}, \penalty0 083011,  (2008).
\newblock \doi{10.1103/PhysRevD.78.083011}.

\bibitem{Kowalski:2014zda}
M.~Kowalski, {Status of High-Energy Neutrino Astronomy}.  (2014).

\bibitem{Ahlers:2014ioa}
M.~Ahlers and F.~Halzen, {Pinpointing Extragalactic Neutrino Sources in Light
  of Recent IceCube Observations}, \emph{Phys.Rev.} {\bf D90}\penalty0 (4),
  \penalty0 043005,  (2014).
\newblock \doi{10.1103/PhysRevD.90.043005}.

\bibitem{Auffenberg:2014hqa}
J.~Auffenberg, {IceVeto: Extended PeV neutrino astronomy in the Southern
  Hemisphere with IceCube}, \emph{AIP Conf. Proc.} {\bf 1630}, \penalty0
  50--53,  (2014).
\newblock \doi{10.1063/1.4902769}.

\bibitem{Schonert:2008is}
S.~Schonert, T.~K. Gaisser, E.~Resconi, and O.~Schulz, {Vetoing atmospheric
  neutrinos in a high energy neutrino telescope}, \emph{Phys.Rev.} {\bf D79},
  \penalty0 043009,  (2009).
\newblock \doi{10.1103/PhysRevD.79.043009}.

\bibitem{Learned:1994wg}
J.~G. Learned and S.~Pakvasa, {Detecting tau-neutrino oscillations at PeV
  energies}, \emph{Astropart.Phys.} {\bf 3}, \penalty0 267--274,  (1995).
\newblock \doi{10.1016/0927-6505(94)00043-3}.

\bibitem{Aartsen:2013jdh}
M.~Aartsen et~al., {Evidence for High-Energy Extraterrestrial Neutrinos at the
  IceCube Detector}, \emph{Science}. {\bf 342}, \penalty0 1242856,  (2013).
\newblock \doi{10.1126/science.1242856}.

\bibitem{Aartsen:2014gkd}
M.~Aartsen et~al., {Observation of High-Energy Astrophysical Neutrinos in Three
  Years of IceCube Data}, \emph{Phys.Rev.Lett.} {\bf 113}, \penalty0 101101,
  (2014).
\newblock \doi{10.1103/PhysRevLett.113.101101}.

\bibitem{Enberg:2008te}
R.~Enberg, M.~H. Reno, and I.~Sarcevic, {Prompt neutrino fluxes from
  atmospheric charm}, \emph{Phys. Rev.} {\bf D78}, \penalty0 043005,  (2008).
\newblock \doi{10.1103/PhysRevD.78.043005}.

\bibitem{ICRC15HESE4}
C.~Kopper, N.~Kurahashi, et~al.
\newblock {Observation of Astrophysical Neutrinos in Four Years of IceCube Data
  }.
\newblock In \emph{Proceeding of the 34th International Cosmic Ray Conference,
  The Hague, Netherlands, 30 July -6 August 2015},  (2015).

\bibitem{Aartsen:2014muf}
M.~Aartsen et~al., {Atmospheric and astrophysical neutrinos above 1 TeV
  interacting in IceCube}, \emph{Phys.Rev.} {\bf D91}\penalty0 (2), \penalty0
  022001,  (2015).
\newblock \doi{10.1103/PhysRevD.91.022001}.

\bibitem{Aartsen:2013vja}
M.~Aartsen et~al., {Energy Reconstruction Methods in the IceCube Neutrino
  Telescope}, \emph{JINST}. {\bf 9}, \penalty0 P03009,  (2014).
\newblock \doi{10.1088/1748-0221/9/03/P03009}.

\bibitem{Aartsen:2015xup}
M.~G. Aartsen et~al., {Measurement of the Atmospheric $\nu_e$ Spectrum with
  IceCube}, \emph{Phys. Rev.} {\bf D91}, \penalty0 122004,  (2015).
\newblock \doi{10.1103/PhysRevD.91.122004}.

\bibitem{ICRC15CASC}
M.~Lesiak-Bzdak, H.~Niederhausen, A.~St\"ossl, et~al.
\newblock {High energy astrophysical neutrino flux characteristics for
  neutrino-induced cascades using IC79 and IC86-string IceCube configurations}.
\newblock In \emph{Proceeding of the 34th International Cosmic Ray Conference,
  The Hague, Netherlands, 30 July -6 August 2015},  (2015).

\bibitem{Aartsen:2015rwa}
M.~G. Aartsen et~al., {Evidence for Astrophysical Muon Neutrinos from the
  Northern Sky with IceCube}, \emph{Phys. Rev. Lett.} {\bf 115}\penalty0 (8),
  \penalty0 081102,  (2015).
\newblock \doi{10.1103/PhysRevLett.115.081102}.

\bibitem{ICRC15LEIF}
S.~Schoenen, L.~R\"adel, et~al.
\newblock {A measurement of the diffuse astrophysical muon neutrino flux using
  multiple years of IceCube data}.
\newblock In \emph{Proceeding of the 34th International Cosmic Ray Conference,
  The Hague, Netherlands, 30 July -6 August 2015},  (2015).

\bibitem{schoenen2015detection}
S.~Schoenen and L.~Raedel, Detection of a multi-pev neutrino-induced muon event
  from the northern sky with icecube, \emph{The Astronomer's Telegram}. {\bf
  7856}, \penalty0 1,  (2015).
\newblock URL \url{http://www.astronomerstelegram.org/?read=7856}.

\bibitem{Acero:2015gva}
F.~Acero, {Fermi Large Area Telescope Third Source Catalog}, \emph{Astrophys.
  J. Suppl.} {\bf 218}\penalty0 (2), \penalty0 23,  (2015).
\newblock \doi{10.1088/0067-0049/218/2/23}.

\bibitem{Wakely:2007qpa}
S.~P. Wakely and D.~Horan.
\newblock {TeVCat: An online catalog for Very High Energy Gamma-Ray Astronomy}.
\newblock In \emph{{Proceedings, 30th International Cosmic Ray Conference (ICRC
  2007)}}, vol.~3, pp. 1341--1344,  (2007).
\newblock URL
  \url{http://indico.nucleares.unam.mx/contributionDisplay.py?contribId=378&confId=4}.

\bibitem{Aartsen:2013bka}
M.~Aartsen et~al., {First observation of PeV-energy neutrinos with IceCube},
  \emph{Phys.Rev.Lett.} {\bf 111}, \penalty0 021103,  (2013).
\newblock \doi{10.1103/PhysRevLett.111.021103}.

\bibitem{AYAICRC}
A.~Ishihara et~al.
\newblock {A search for extremely high energy neutrinos in 6 years of IceCube
  data}.
\newblock In \emph{Proceeding of the 34th International Cosmic Ray Conference,
  The Hague, Netherlands, 30 July -6 August 2015},  (2015).

\bibitem{Beresinsky:1969qj}
V.~S. Berezinsky and G.~T. Zatsepin, {Cosmic rays at ultrahigh-energies
  (neutrino?)}, \emph{Phys. Lett.} {\bf B28}, \penalty0 423--424,  (1969).
\newblock \doi{10.1016/0370-2693(69)90341-4}.

\bibitem{Beacom:2001xn}
J.~F. Beacom, P.~Crotty, and E.~W. Kolb, {Enhanced signal of astrophysical tau
  neutrinos propagating through earth}, \emph{Phys. Rev.} {\bf D66}, \penalty0
  021302,  (2002).
\newblock \doi{10.1103/PhysRevD.66.021302}.

\bibitem{Gondolo:1995fq}
P.~Gondolo, G.~Ingelman, and M.~Thunman, {Charm production and high-energy
  atmospheric muon and neutrino fluxes}, \emph{Astropart. Phys.} {\bf 5},
  \penalty0 309--332,  (1996).
\newblock \doi{10.1016/0927-6505(96)00033-3}.

\bibitem{hallenmeasurement}
P.~Hallen.
\newblock \emph{On the Measurement of High-Energy Tau Neutrinos with IceCube}.
\newblock PhD thesis, Master's thesis, RWTH Aachen, November 2013. URL
  https://internal. icecube. wisc. edu/reports/details. php.

\bibitem{ICRC15TAU}
D.~R. Williams, C.~M. Vraeghe, D.~L. Xu, et~al.
\newblock {A Search for Astrophysical Tau Neutrinos in Three Years of IceCube
  Data}.
\newblock In \emph{Proceeding of the 34th International Cosmic Ray Conference,
  The Hague, Netherlands, 30 July -6 August 2015},  (2015).

\bibitem{HANSSEBINT}
H.~Niederhausen and S.~Schoenen.
\newblock Xxxxxx.
\newblock IceCube Internal Report ICECUBE-XXXX2015, Department Physics and
  Astronomy, Stony Brook and III.Physikalisches Institut RWTH Aachen
  University,  (2015).
\newblock available at \url{http://.pdf}.

\bibitem{Aartsen:2015knd}
M.~G. Aartsen et~al., {A combined maximum-likelihood analysis of the
  high-energy astrophysical neutrino flux measured with IceCube},
  \emph{Astrophys. J.} {\bf 809}\penalty0 (1), \penalty0 98,  (2015).
\newblock \doi{10.1088/0004-637X/809/1/98}.

\bibitem{Aartsen:2013jla}
M.~G. Aartsen et~al.
\newblock {The IceCube Neutrino Observatory Part II: Atmospheric and Diffuse
  UHE Neutrino Searches of All Flavors}.
\newblock In \emph{{Proceedings, 33rd International Cosmic Ray Conference
  (ICRC2013)}},  (2013).
\newblock URL
  \url{https://inspirehep.net/record/1255627/files/arXiv:1309.7003.pdf}.

\bibitem{Aartsen:2014cva}
M.~G. Aartsen et~al., {Searches for Extended and Point-like Neutrino Sources
  with Four Years of IceCube Data}, \emph{Astrophys. J.} {\bf 796}\penalty0
  (2), \penalty0 109,  (2014).
\newblock \doi{10.1088/0004-637X/796/2/109}.

\bibitem{Aartsen:2014ivk}
M.~G. Aartsen et~al., {Searches for small-scale anisotropies from neutrino
  point sources with three years of IceCube data}, \emph{Astropart. Phys.} {\bf
  66}, \penalty0 39--52,  (2015).
\newblock \doi{10.1016/j.astropartphys.2015.01.001}.

\bibitem{Margiotta:2014gza}
A.~Margiotta, {The KM3NeT deep-sea neutrino telescope}, \emph{Nucl. Instrum.
  Meth.} {\bf A766}, \penalty0 83--87,  (2014).
\newblock \doi{10.1016/j.nima.2014.05.090}.

\bibitem{Avrorin:2013sla}
A.~V. Avrorin et~al., {Current status of the BAIKAL-GVD project}, \emph{Nucl.
  Instrum. Meth.} {\bf A725}, \penalty0 23--26,  (2013).
\newblock \doi{10.1016/j.nima.2012.11.151}.

\bibitem{Aartsen:2014njl}
M.~Aartsen et~al., {IceCube-Gen2: A Vision for the Future of Neutrino Astronomy
  in Antarctica}.  (2014).

\end{thebibliography}
\blankpage
%\printindex[aindx]                 % to print author index
\printindex                         % to print subject index
\end{document}